# Magnetothermopower and Magnetoresistivity of $RuSr_2Gd_{1-x}La_xCu_2O_8$ (x=0, 0.1)

C.-J. Liu,[1] C.-S. Sheu,[1] T.-W. Wu,[1] L.-C. Huang,[1] F. H. Hsu,[2] H. D.Yang,[2] G. V. M. Williams[3], Chia-Jung C. Liu[4]

[1]Department of Physics, National Changhua University of Education, Changhua 500, Taiwan, R. O. C.

[2]Department of Physics, National Sun Yat-Sen University, Kaohsiung 804, Taiwan, R. O. C.

[3]The New Zealand Institute for Industrial Research, P. O. Box 31310, Lower Hutt, New Zealand

[4]Department of Chemistry, Medical Technology, and Physics, Monmouth University, West Long Branch, New Jersey 07764-1898, U. S. A.

We report measurements of magnetothermopower and magnetoresistivity as a function of temperature on $RuSr_2Gd_{1-x}La_xCu_2O_8$ (x = 0, 0.1). The normal-state thermopower shows a dramatic decrease after applying a magnetic field of 5 T, whereas the resistivity shows only a small change after applying the same field. Our results suggest that $RuO_2$ layers are conducting and the magnetic field induced decrease of the overall thermopower is caused by the decrease of partial thermopower associated with the spin entropy decrease of the carriers in the $RuO_2$ layers.

The coexistence of superconductivity and magnetic ordering in the hybrid ruthenate-cuprates $RuSr_2GdCu_2O_8$ (Ru-1212) has attracted considerable attention because of the antagonism between superconductivity and ferromagnetism.[1-8] Low field antiferromagnetic order is observed in Ru-1212 and there is a spin-flop transition to ferromagnetic order as the magnetic field is increased.[4,9] The magnetic ordering temperatures vary between 138 K and 132 K,[4,6] while superconductivity arising from the $CuO_2$ layers is observed below 46 K. Bulk superconductivity in Ru-1212 was reported by specific-heat measurements[5] and observation of a bulk Meissner signal in dc magnetization measurements.[6] On the contrary, based on the reversible magnetization measurements,[10] the results indicate the absence of a Meissner state. A Josephson-junction-array model is proposed to explain the observed data in Ru-1212 that superconducting transition temperature is largely suppressed by the magnetic field with the decrease rate of $dT_c/dH \approx 100$ K/T observed and the extremely field-sensitive diamagnetic effect due to the large intragrain penetration depth.[11]

These compounds bear a resemblance to the structures of $NbBa_2RCu_2O_8$[12], and HTSC compounds. In the four number naming scheme, Ru-1212 is viewed as in the same category as $CuBa_2YCu_2O_{7-\delta}$ (Y-123) but with the $RuO_2$ layers replacing the CuO chains. The conducting nature of CuO chains in Y-123 has been confirmed by the measurements of electrical conductivity,[13,14] positron annihilation,[15] and thermoelectric power (TEP).[16] One of the Bi-O bands along the (100) direction dipping below $E_F$ derived from the linearized augmented-plane-wave method and the mixed valence of Bi in Bi-based HTSC from a XANES study both suggest the contribution of $BiO_{1+\delta}$ layers to the electronic transport.[17,18] It has also been suggested that the electron carriers in the $BiO_{1+\delta}$ layers are responsible for the negative TEP observed at high temperatures in Bi-based HTSC.[19] The evidence of mixed valence of Ru in Ru-1212 has been provided by NMR[20,21] and XANES studies.[22] Moreover, in a microscopic t-J-I model study, that takes into account the antiferromagnetic (J) and ferromagnetic (I) exchange interactions simultaneously, it has been suggested that the hole carriers are located at the lower Hubbard band (LHB) top for $CuO_2$ layers and the electron carriers located at the upper Hubbard band (UHB) bottom for the $RuO_2$ layers.[23]

TEP measurements can provide both information of the type and the characteristic energy of charge carriers and therefore are a complementary tool to the resistivity measurements for studying the transport properties. Since TEP is a measure of the heat per carrier over temperature, we can thus view it as a measure of the entropy per carrier. Both $La_{1.85}Sr_{0.15}CuO_{4-\delta}$ and $YBa_2Cu_3O_{7-\delta}$ show a field-independent of TEP up to 30 T, indicating retaining no spin degree of freedom in the $CuO_2$ layers.[24] The $RuO_2$ layers in the magnetic superconductors are known to be responsible for the magnetic order. We are therefore motivated to examine the TEP of Ru-1212 measured in a magnetic field to understand further the role of $RuO_2$ layers in the transport properties. In this paper, we present the thermopower and electrical resistivity as a function of temperature in zero field and 5 T for La-doped Ru-1212 and find small magnetoresistance (MR) effects but large magnetothermopower effects in both samples.

The $RuSr_2Gd_{1-x}La_xCu_2O_8$ (x = 0, 0.1) ceramics were synthesized by quantitatively mixing high-purity powders of $RuO_2$, $SrCO_3$, $Gd_2O_3$, CuO, and $La_2O_3$. Note that the x = 0.1 composition for the La-doped Ru-1212 is near the solubility limit of La.[25] The mixed powders were calcined at 960℃ in air for 12 h, followed by sintering at 1010℃ in flowing nitrogen for 10 h and then 1055℃ in flowing oxygen for 10 h. The sintering in nitrogen is essential to suppress the formation of $SrRuO_3$. Finally, the samples were annealed at 1060℃ in flowing $O_2$ for 7 days. The resulting samples were examined by a powder x-ray diffractometer equipped with Co $K_\alpha$ radiation. The x-ray diffraction patterns indicate no evidence of the presence of $SrRuO_3$ impurity within the x-ray detection limit. Electrical resistivity as a function of temperature was measured using standard dc four-probe techniques. Thermopower as a function of temperature was measured using steady state techniques. A Cernox thermometer was used to monitor the temperature of the sample. The temperature gradient across the sample was monitored by two chromel-constantan thermocouples connected in a differential mode. The thermally generated Seebeck voltage across the sample was measured using a Keithley nanovoltmeter. Thermopower data were subtracted from the Seebeck probe Cu leads (25 μm diameter, Puratronic from Alfa Aesar, 99.995%). The thermopower of Cu leads were calibrated against Pb (99.9995%) by comparing with Robert's data[34] at $T \geq 90$ K and a HTSC $YBa_2Cu_3O_7$. at $T < 90$ K. We realize that magnetothermopower of the Seebeck probe leads would have background contribution at low temperatures and in high magnetic fields. However, this would not affect our discussion on thermopower data, since the magnetothermopower effect is much more evident at high T than that at low T. A commercial SQUID magnetometer (Quantum Design) was used to provide the magnetic field with the direction parallel to the temperature gradient or electric current for transport measurements.

Fig. 1 shows the mass susceptibility of superconducting (SC) Ru-1212 (x = 0) and nonsuperconducting (non-SC) Ru-1212 (x = 0.1). A pronounced peak of susceptibility is observed at the magnetic ordering temperature $T_m \approx 132$ K and the diamagnetic transition occurs near 20 K for SC Ru-1212. The magnetic ordering temperature is shifted to a higher temperature at $T_m \approx 155$ K and no diamagnetic transition is observed for non-SC Ru-1212. Note that the magnetic ordering temperature also shifts to high temperatures for $Ru(Sr_{1-x}La_x)_2GdCu_2O_8$, which has been ascribed to the enhanced superexchange interaction in the Ru sublattice induced by the La substitution.[26] As shown in Fig. 2, the resistivity data show the occurrence of superconductivity at $T_{c, zero}$ = 28 K in zero field and $T_{c, zero}$ = 7 K in 5 T for SC Ru-1212 (x = 0). Note that the temperature dependence of resistivity and superconducting transition temperature strongly depends on the preparation conditions. The nonmetallic temperature dependence of resistivity of Ru-1212 in ref.1 was ascribed to the oxygen deficiency.[1] Nevertheless, the Rietveld refinement of Ru-1212 structures reveals that the oxygen annealing could cause the variation of the cation composition on the 10-fold coordinated *2h* site of Sr, the prismatic 8-fold coordinated *1d* site of Gd site and the *4l* site of Ru without changing the oxygen content.[27] The intermixing of Gd and Sr or the Ru and Cu leads to the variation of the formal valence of Cu, which in turn is expected to affect the transport properties. Fig. 3 shows the temperature dependence of resistivity for non-SC Ru-1212 (x = 0.1). No superconducting transition is observed down to 5 K. The magnetoresistance is evident for both samples of SC and non-SC Ru-1212. The MR ratio, defined as [R(0T)-R(5T)]/R(5T), is small with the ratio less than 4%. The magnetoresistance could be ascribed to spin scattering contribution in the magnetic $RuO_2$ layers since the hole carriers in the $CuO_2$ layers are supposed to retain no spin degree of freedom like other HTSC compounds.

For materials with more than one type of charge carrier, the diffusion thermopower can be expressed as

$$S = \sum_i \left( \frac{\sigma_i}{\sigma} \right) S_i , \qquad (1)$$

where $\sigma_i$ and $S_i$ are respectively the electrical conductivity and partial thermopower associated with the ith group of carriers. If the charge carriers retain a spin degree of freedom, a spin entropy term, $(k_B/q)ln2$, would contribute to the TEP in the absence of a magnetic field.[24] Once the spin is forced to align with a large magnetic field, the spin entropy would decrease towards zero. In Fig. 4, we plot the TEP versus temperature in zero field and H = 5 T for SC (x = 0) and non-SC (x = 0.1) $RuSr_2Gd_{1-x}La_xCu_2O_8$. The TEP data exhibit several features as described in the following. Firstly, the sign of charge carriers is positive for both SC and non-SC Ru-1212, indicating the majority carriers are holes according to Eq. (1). Secondly, the TEP of the title samples exhibits a broad hump behavior, a common characteristic of underdoped high-$T_c$ systems, and decreases with decreasing temperature in the low temperature regime. These results are consistent with the Hall effect measurements by McCrone et al.[32] The Hall effect data bear similarities to those for underdoped $YBa_2Cu_3O_{7-\delta}$[33] in the temperature region above $T_m$ except a downturn for the SC Ru-1212 below 160 K. The downturn might be due to the itinerant character of the charge carriers in the $RuO_2$ layers below $T_m$. Besides, the Hall coefficient of SC Ru-1212 is positive, which confirms its hole character of the majority carriers. According to the Mott formula,[28] the diffusion TEP of a metal is given as

$$S_D = \frac{\pi^2}{3}\frac{k_B}{e}k_B T\left(\frac{\partial \ln \sigma(E)}{\partial E}\right)_{E=E_F}, \quad (2)$$

where σ(E) is a conductivity-like function for carries of energy E measured from the Fermi level, *e* the carrier charge (negative for electrons and positive for holes), and $k_B$ the Boltzmann constant. Therefore, the characteristic diffusion TEP of a metal decreases with decreasing temperature and is small. Nevertheless, this is not the case for the title samples because of their nonmetallic temperature dependence of resistivity. For variable-range-hopping transport in a disordered system, the temperature dependence of resistivity and TEP should respectively follow the forms[29]

$$\sigma = \sigma_0 \exp\left[-\left(\frac{T_0}{T}\right)^{\frac{1}{d+1}}\right] \quad (3)$$

and

$$S(T) \propto T^{\frac{d-1}{d+1}}, \tag{4}$$

where $\sigma_0$ is weakly temperature dependent, $T_0$ associated with the localization length, $d$ the dimensionality. For 3D, the conductivity $ln\ \sigma$ should vary as $T^{-1/4}$ and the TEP should vary as $T^{1/2}$. In Fig. 5 we replot the TEP in zero field of SC Ru-1212 against $T^{1/2}$ and find that a good linear dependence is observed at $50 \leq T \leq 160$ K. In the inset of Fig. 5, it can be seen that the conductivity follows the Mott's $T^{1/4}$ law quite nicely $52 \leq T \leq 106$ K. Similar behavior of resistivity and TEP for non-SC Ru-1212 is observed. Together with the fact that the intermixing of Gd and Sr or the Ru and Cu leads to the variation of the formal valence of Cu, it seems to suggest that the disorder of cations plays a role at low T in the electronic transport of the title samples, reminiscent of the case in non-SC $La_{1.8}Sr_{0.2}CaCu_2O_{6-\delta}$ following the Mott's $T^{1/4}$ law in conductivity and $T^{1/2}$-law in TEP at low T.[30] In $(La,Sr,Ca)_3Cu_2O_{6-\delta}$, the $O_2$ annealing could change the cation ratio at both the *2a* and *4e* sites, which turns out to have significant effects on their transport properties. On the other hand, the disorder might come from the granularity in terms of microstructural inhomogeneity of Ru-1212. In a study by the high-resolution electron microscopy (HRTEM) and synchrotron x-ray-diffraction methods,[31] the HRTEM image shows that there exists the disorder in the average structure and microstructure based on the observations of many small misoriented domains in as-prepared Ru-1212 due to the near coincidence of lattice constants *a* and *b* with *c/3*, and the subdomains separated by sharp antiphase boundaries in a well-annealed sample due to the rotations of the $RuO_6$ octahedra around *c* axis. High-temperature annealing process would help improve the domain microstructure but not completely. As compared to the metal-like temperature dependence of the resistivity in the high temperature regime in ref. 5, our nonmetallic temperature dependence seems to be influenced more by the inhomogeneity of microstructure even though with similar preparation conditions. In addition, disorder also exists in the oxygen atoms in the $RuO_2$ planes and of the apical oxygen atoms linking the $CuO_5$ units and $RuO_6$ octahedra evidenced by the large atomic displacement *U* factors.

Thirdly, the TEP from both SC and non-SC samples strikingly shows a significant decrease after applying a field of 5 T when considering the small change in resistivity with the same field applied. This can not be due to the $CuO_2$ layers, because it has been found that there is no magnetic field dependence of TEP for $La_{1.85}Sr_{0.15}CuO_4$ and $YBa_2Cu_3O_{7-\delta}$ up to 30 T,[24] instead it should be due to the contribution from the conducting $RuO_2$ layers. A number of studies have shown that the magnetic order is attributed to the Ru moment in the $RuO_2$ layers and the magnetic order above ~ 4 kG is ferromagnetic.[9] In addition, based on calculations using the local density approximation with a Coulomb repulsion $U_{Ru}$ = 3 eV (LDA+U) and generalized gradient approximation (GGA) procedures, the $RuO_2$ layers from Ru-1212 are predicted to be metallic,[3] which has been suggested by a Hall effect and magnetoresistance study.[8] Hence, it is plausible that the TEP of Ru-1212 arises from transport in both the $CuO_2$ and $RuO_2$ layers according to Eq. (1). We can rewrite Eq. (1) as

$$S = \frac{\sigma_{CuO_2}}{\sigma_{CuO_2} + \sigma_{RuO_2}} S_{CuO_2} + \frac{\sigma_{RuO_2}}{\sigma_{CuO_2} + \sigma_{RuO_2}} S_{RuO_2} . \qquad (5)$$

In view of the small magnetoresistance, the decrease of TEP induced by the magnetic field has relatively small contribution from the weighting factor of conductivity in Eq. (5) and therefore is mainly from the decrease of partial TEP of $S_{RuO_2}$ in the $RuO_2$ layers. As the spins on the Ru sites are aligned with the field, the decrease in the spin entropy could lead to a concomitant decrease in TEP of $S_{RuO_2}$. This is evidenced by the fact that TEP becomes less field dependent at low T as long-range ferromagnetism develops. Following this reasoning, the temperature-dependent magnetothermopower should follow a function like (1-f), where f is an order parameter for the ferromagnetic order. In Fig. 6, we plot the magnetothermopower, S(0Oe)-S(5T), from the data in Fig. 4 and the magnetization measured at 5 T as a function of temperature. As the ferromagnetic order develops (the magnetization increases) at low T, the magnetothermopower becomes smaller. The magnetothermopower ratio, defined as [S(0T)-S(5T)]/S(5T), is estimated to be 28% at 190 K for SC Ru-1212 (x = 0), 117% at 240 K for non-SC Ru-1212 (x = 0.1), respectively.

In summary, we have measured TEP and resistivity as a function of temperature on $RuSr_2Gd_{1-x}La_xCu_2O_8$ (x = 0, 0.1) in zero field and H = 5 T. At H = 5 T, there is a significant decrease in TEP for both SC (x = 0)and non-SC (x = 0.1) samples that could be ascribed to the decrease in the spin entropy contribution associated with the carriers in the $RuO_2$ layers, as the Ru moments are aligned with the magnetic field. The conductivity of both samples at low T follows the form of $T^{-1/4}$ and the TEP follows the form of $T^{1/2}$, suggesting disorder plays a role at low T in the present Ru-1212 system.

**ACKNOWLEDGMENTS**

This research is supported by National Science Council of ROC under grant Nos. NSC90-2112-M-018-006, NSC-92-2112-M110-017 and the New Zealand Marsden Fund.

Figure Captions

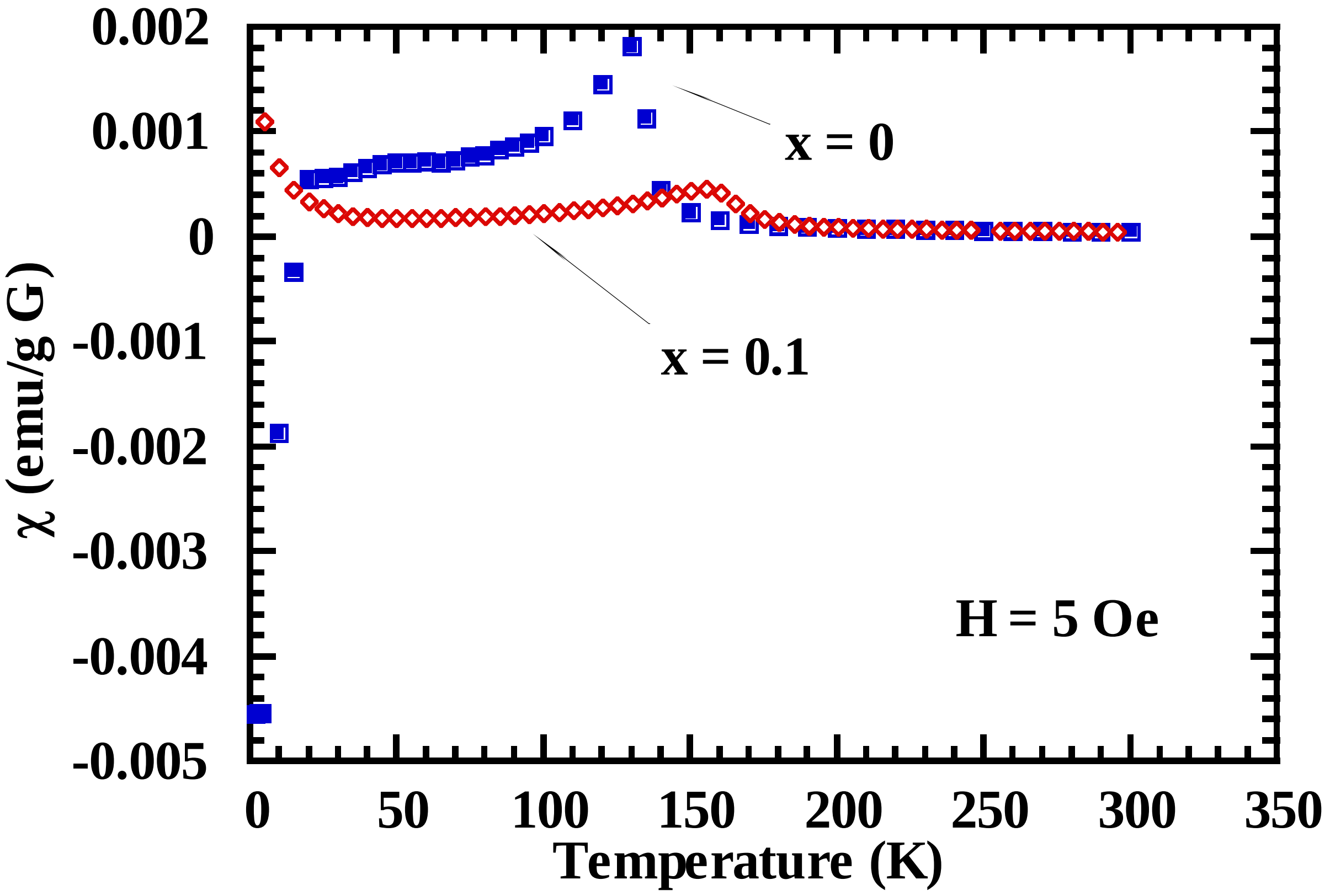

Fig. 1. Temperature dependence of the zero-field-cooled dc mass susceptibility of $RuSr_2GdCu_2O_8$ and $RuSr_2Gd_{0.9}La_{0.1}Cu_2O_8$ measured at 5 Oe.

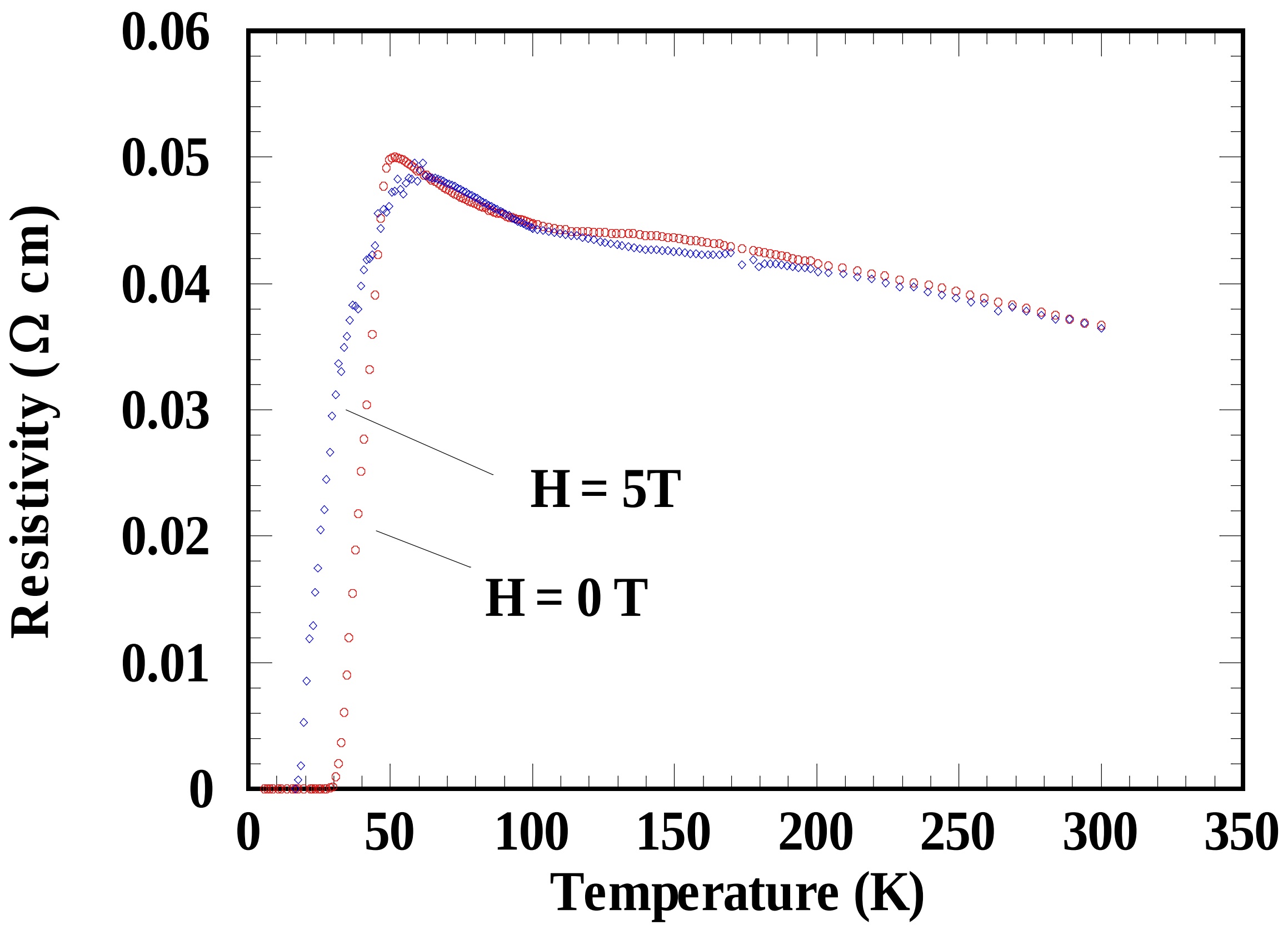

Fig. 2. Temperature dependence of the resistivity in zero field and H = 5 T for $RuSr_2GdCu_2O_8$.

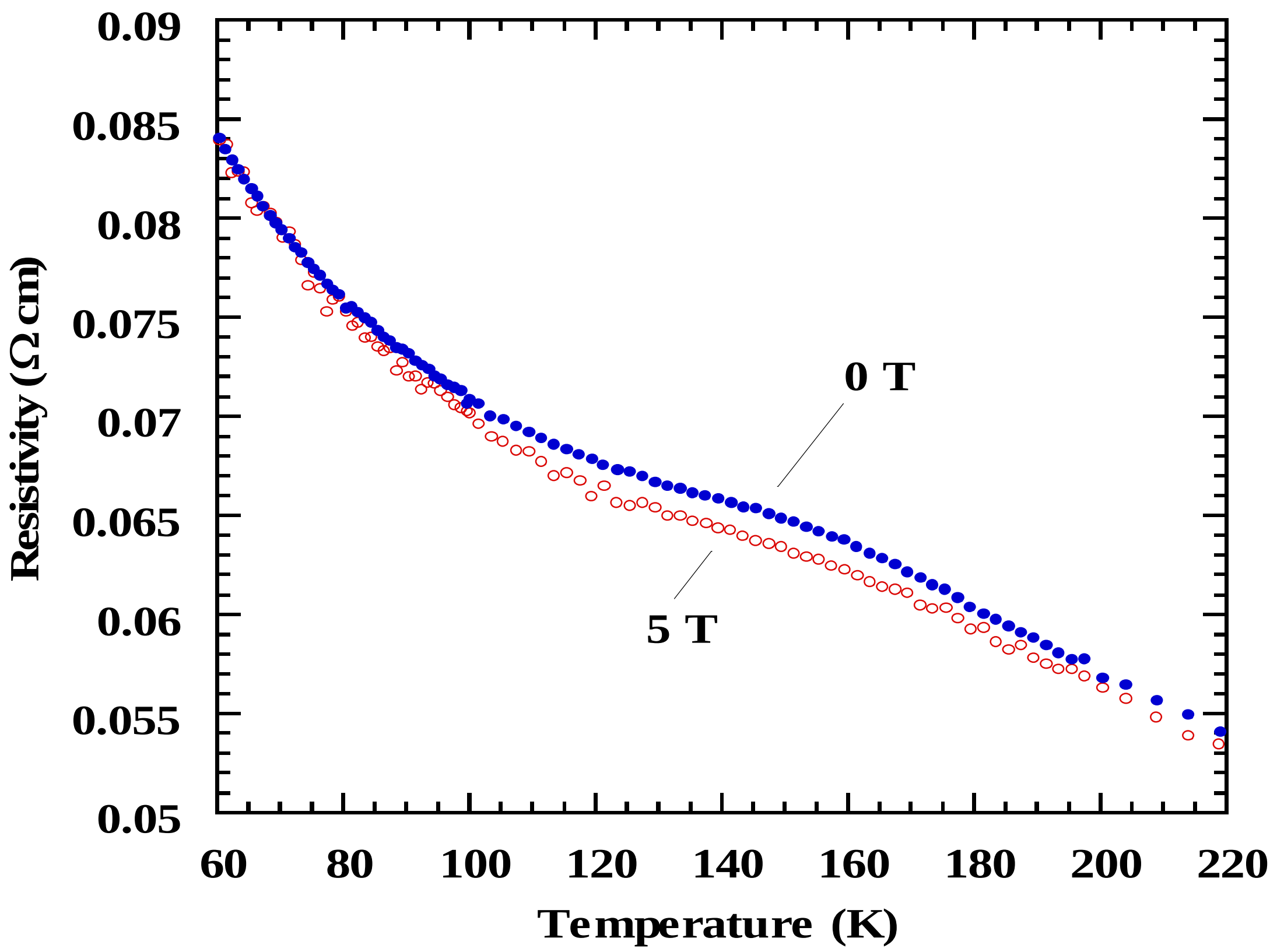

Fig. 3. Temperature dependence of the resistivity in zero field and H = 5 T for $RuSr_2Gd_{0.9}La_{0.1}Cu_2O_8$. In order to have the magnetoresistance visible, the whole data set is not shown. This sample shows no superconducting transition down to 5 K.

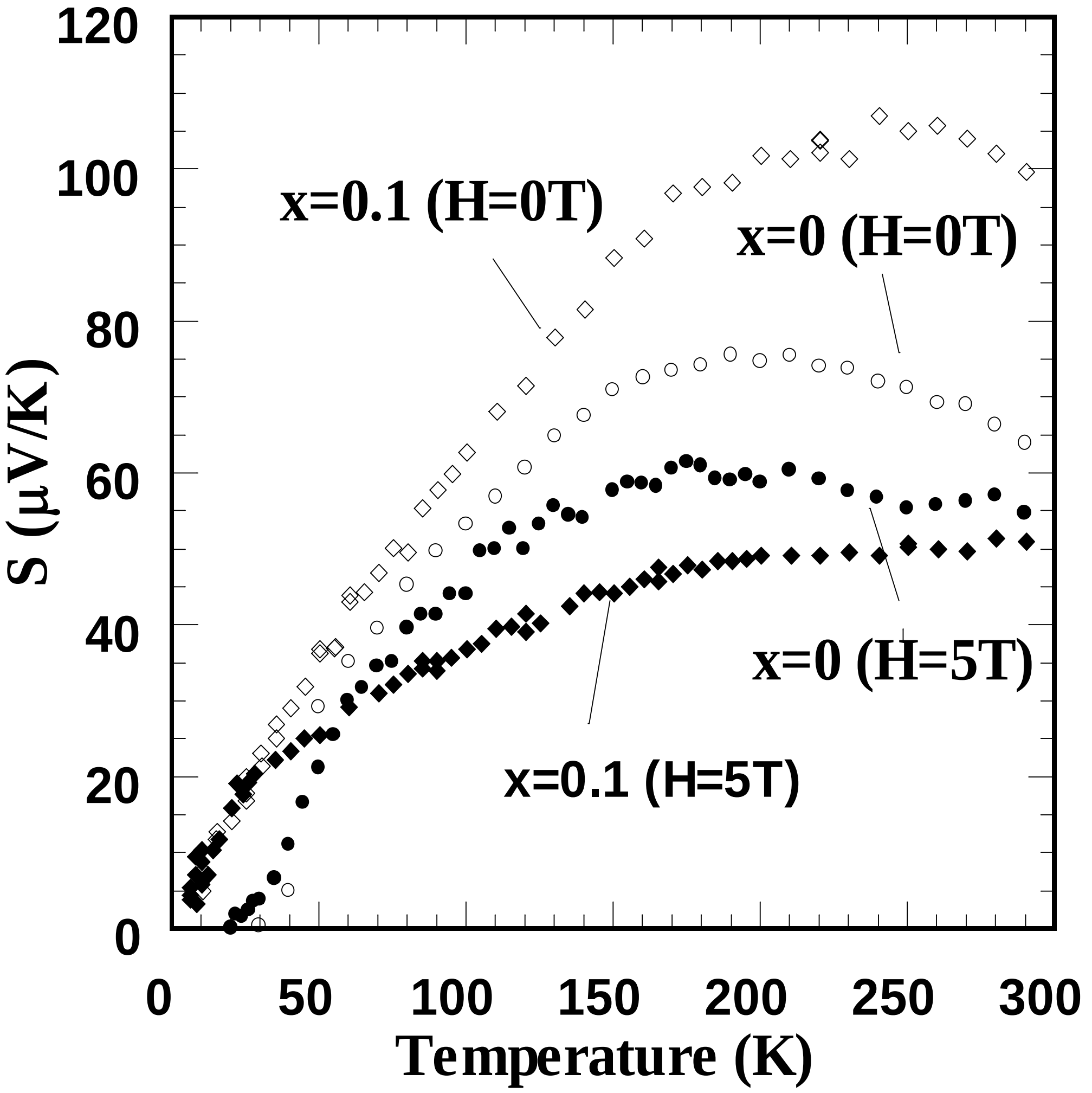

Fig. 4. Temperature dependence of thermopower in zero field and H = 5 T for $RuSr_2Gd_{1-x}La_xCu_2O_8$ (x = 0, 0.1).

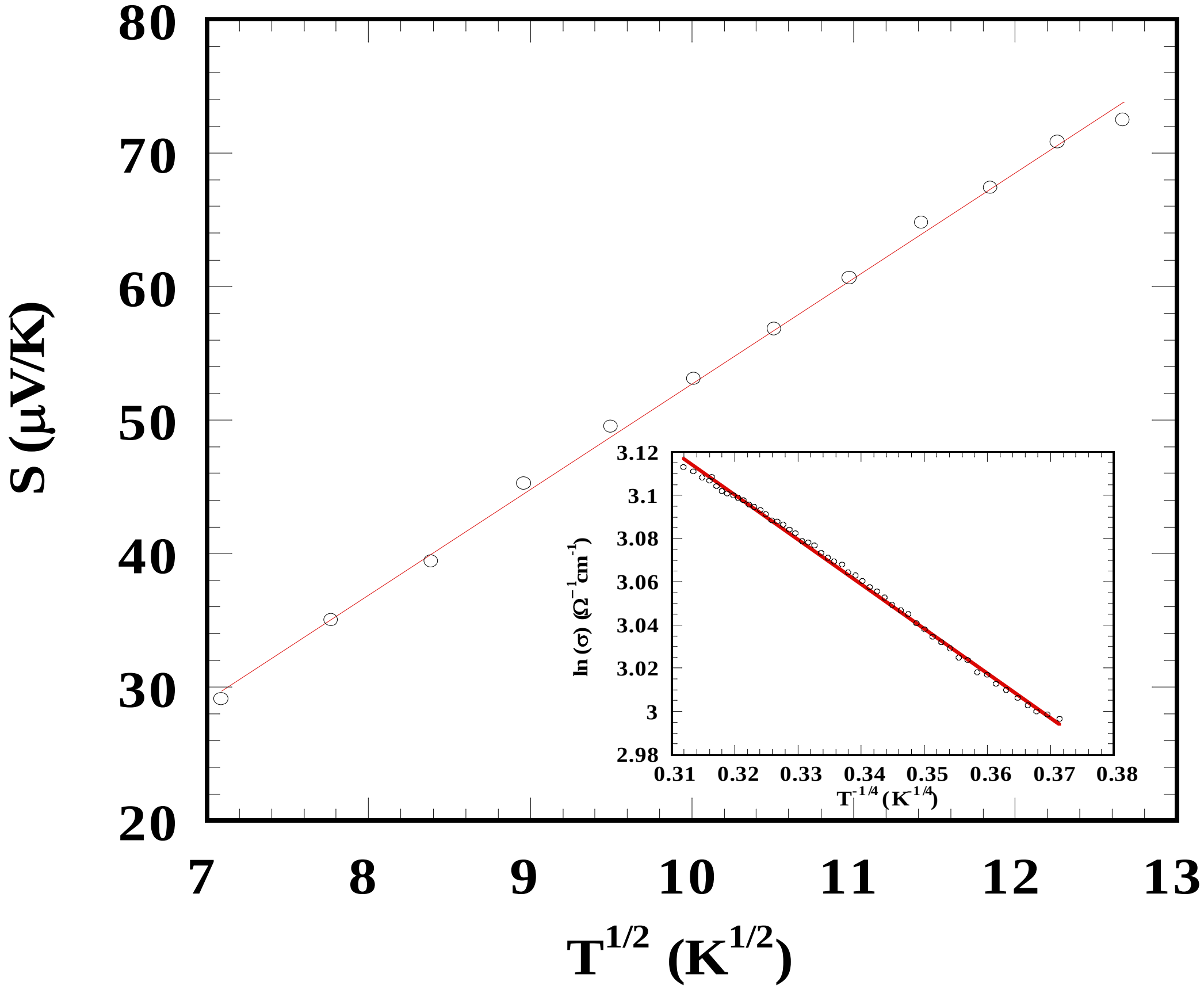

Fig. 5. TEP for SC Ru-1212 at $50 \leq T \leq 160$ K varies as $T^{1/2}$, suggestive a variable range hopping process in 3D. The inset illustrates the Mott's $T^{1/4}$-law of conductivity in a variable range hopping process.

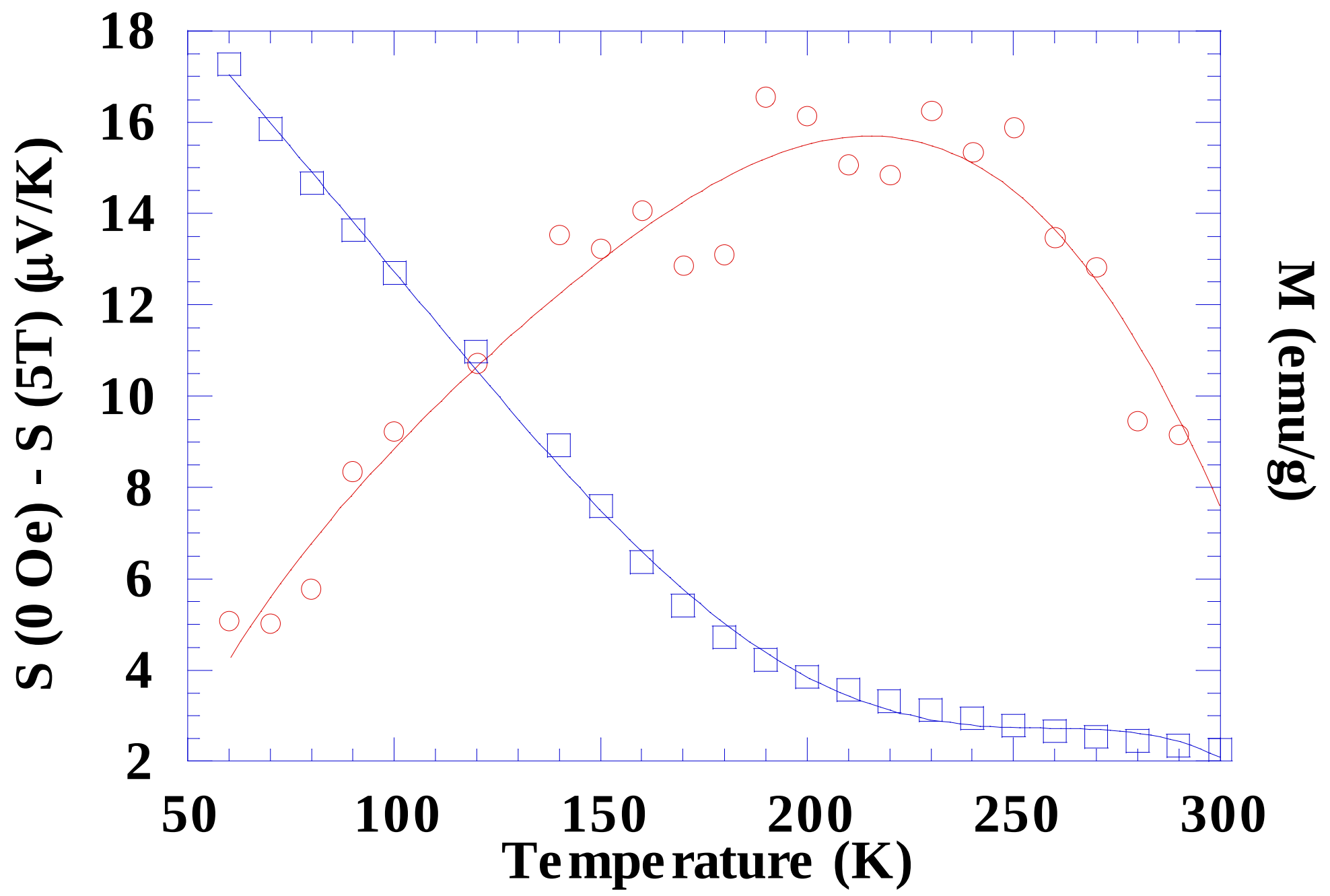

Fig. 6. The magnetothermopower, S(0Oe)-S(5T), from the data in Fig. 4 and the magnetization measured at 5 T as a function of temperature. As the ferromagnetic order develops (the magnetization increases) at low T, the magnetothermopower becomes smaller.